\documentclass[twocolumn]{article}
\usepackage{graphicx}

\begin{document}

\title{Mechanism for Surface Waves in \\
Vibrated Granular Material}

\author{Jysoo Lee \\
\\
Department of Physics, Seoul National University \\
Seoul 151-742, Korea \\
and \\
Benjamin Levich Institute and Department of Physics \\
City College of the City University of New York \\
New York, NY 10031}

\date{March 2, 1999}

\maketitle

\begin{abstract}

We use molecular dynamics simulations to study the formation of
surface waves in vertically vibrated granular material.  We find that
horizontal movements of particles, which are essential for the
formation of the waves, consist of two distinct processes.  First, the
movements sharply increase while the particles are colliding with a
bottom plate, where the duration of the collisions is very short
compared to the period of the vibration.  Next, the movements
gradually decrease between the collisions, during which the particles
move through the material.  We also find that the horizontal velocity
field after the collisions is strongly correlated to the surface
profile before the collisions.


\end{abstract}

\section*{}


Granular material under vibration has been a constant source of
interesting phenomena among which surface waves are particularly
interesting \cite{c90,hb92,mb94,hnst95,jnb96}.  Particles in a
vertically vibrating box display a variety of steady state
patterns---stripes, squares, hexagons, and localized excitations
called ``oscillons'', and even a richer set of transient patterns
\cite{fdl89,mus94,mus95,mus96,mkj97,ums98}.  The fact that such
diverse patterns can exist in the seemingly simple system has
attracted a lot of efforts on the study of the mechanism for the
formation of the surface waves
\cite{cmr97,s97,er98,ta97,atv98,sb97,r98,vo98}.

It is generally accepted that horizontal movements of particles play
an important role in the formation of the surface waves.  The present
theories can be roughly divided into two groups according to how such
movements are generated and maintained.  In the first group of
theories, it is argued that horizontal movements are generated while
particles are colliding with the bottom of a box
\cite{cmr97,s97,er98}.  The duration of the collisions is very short
compared to the period of the vibration.  It is also argued that
between the collisions with the bottom, horizontal movements gradually
decrease while the particles move through the medium.  Using these
assumptions, many of the patterns observed in the experiments can be
reproduced.  In the other group of theories, it is argued that
horizontal movements depend only on continuum variables such as height
and/or density fields \cite{ta97,atv98,sb97,r98}.  Since these fields
change smoothly over time, changes in the movements are also gradual.
In particular, changes in the movements during the collisions (with
the bottom) are not particularly different from those during other
phases of the vibration.  These theories can also reproduce many of
the experimental patterns, typically by coupling density and height
fields.

The underlying assumptions of all of these theories sound plausible.
Also, it is very difficult to determine that which of the theories
describes the experiments better.  All of them reproduce many of the
experimental patterns, and have their own strong and weak points.
Also, one should note that such theories cannot be validated just
because they reproduce the observed patterns.  Quantitative agreements
(e.g., of dispersion relation) are necessary for the validation.  In
order to check the validity of the theories, it is thus necessary to
obtain detailed information on the system, and compare it with the
predictions of the theories.  Unfortunately, such information is
usually difficult to obtain by experiments.


In this paper, we use molecular dynamics (MD) simulation method, which
provides detailed information on the motion of individual particles as
well as time averaged fields.  We study the system in two dimensions.
We focus on horizontal movements of particles, which are essential to
the formation of the surface waves.  We find that the time evolution
of average horizontal speed $U$ is made up of two separate processes.
First, there are sharp increases in $U$ during the short periods that
the pile is colliding with the bottom plate.  The other process is
gradual decay of $U$ between such collisions, which results from
interparticle collisions.  The time evolution of $U$ strongly supports
the theories in the first group \cite{cmr97,s97,er98}.  The present
results do not imply that a continuum description is not possible for
the system, it implies that the interpretation of the continuum
variables has to be modified.

We then study the processes in more detail.  We find that the
horizontal velocity field $V_{x}(x)$ after collisions (with the
bottom) shows a strong correlation with $\partial _{x} h(x)$ before
the collisions, where $h(x)$ is the center of mass field.  Such
correlation is assumed in the theories of \cite{cmr97,er98}.  The
second process can be characterized by temporal decay of $U$.  We find
that the decay time is very small when there is no surface wave, and
it is comparable to the period of the vibration when surface waves are
present.  We also study the parameter dependence of the decay time and
maximum horizontal speed.



The simulations are done in two dimensions with disk shaped particles,
using a form of interaction due to Cundall and Strack
\cite{cs79,lh93}.  Particles interact only by contact, and the force
between two such particles $i$ and $j$ is the following.  Let the
coordinate of the center of particle $i$ ($j$) be $\vec{R}_i$
($\vec{R}_j$), and $\vec{r} = \vec{R}_i - \vec{R}_j$.  The normal
component $F_{j \to i}^{n}$ of the force acting on particle $i$ from
particle $j$ is
\begin{equation}
\label{eq:fn}
F_{j \to i}^{n} = k_n (a_i + a_j - \vert \vec{r} \vert) 
                - \gamma m_e (\vec{v} \cdot \hat{n}),
\end{equation}
where $a_i$ ($a_j$) is the radius of particle $i$ ($j$), $\hat{n} =
\vec{r} / r$, and $\vec{v} = d\vec{r}/dt$.  Here, $k_n$ is the elastic
constant, $\gamma$ the friction coefficient, and $m_e$ is the
effective mass, $m_i m_j/(m_i + m_j)$.  The shear component $F_{j \to
i}^{s}$ is given by
\begin{equation}
F_{j \to i}^{s} = - {\rm sign} (\delta s) ~ 
{\rm min}(k_s \vert \delta s \vert, \mu \vert F_{j \to i}^n \vert),
\label{eq:fs}
\end{equation}
where $\mu$ is the friction coefficient, $\delta s$ the {\em total}
shear displacement during a contact, and $k_s$ is the elastic constant
of a virtual tangential spring.  The shear force applies a torque to
the particles, which then rotate.

Particles can also interact with walls.  The force and torque on
particle $i$, in contact with a wall, are given by (\ref{eq:fn}) -
(\ref{eq:fs}) with $a_j = 0$ and $m_e = m_i$.  Also, the system is in
a vertical gravitational field $\vec{g}$.  The interaction parameters
used in this study are fixed as follows, unless otherwise specified:
$k_n = k_s = 10^{7}, \gamma = 10^{4}$ and $\mu = 0.2$.  And, the
timestep for integration is $5 \times 10^{-7}$.  In order to avoid
artifacts of a monodisperse system (e.g., hexagonal packing), we
choose the radius of the particles from a Gaussian distribution with
the mean $0.1$ and width $0.02$.  The density of the particles is $5$.
Throughout this paper, CGS units are implied.


We put particles on a horizontal plate which oscillates sinusoidally
along the vertical direction with given amplitude $A$ and frequency
$f$.  Let the width of the plate be $W$.  We apply a periodic boundary
condition in the horizontal direction.  We start the simulation by
inserting the particles at random positions above the plate. We let
them fall by gravity and wait while they lose energy by collisions. We
wait for $10^{6}$ iterations for the particles to relax, and during
this period we keep the plate fixed.  The typical velocity at the end
of the relaxation is of order $10^{-2}$.  After the relaxation, we
vibrate the plate, and start to take measurements.


The coefficient of restitution between the particles $e_{pp}$,
determined from the above interaction parameters, is $0.21$, and the
coefficient between the particles and the plate $e_{pw}$ is $8.0
\times 10^{-2}$.  The particles are thus strongly inelastic.  We have
studied the motion of a single particle for several values of $A$ with
$f = 10$, and find good agreements with the predictions by Mehta and
Luck \cite{ml90}.  Also, the present model was shown to reproduce the
dispersion relation from experiment \cite{whb96}.


We then study the center of mass motion of the particles.  As the
depth of the pile increases, its motion can be different from that of
a single particle \cite{l98}.  For the parameters with which the
motion of a single particle is periodic with period $T = 1/f$, the
motion of the pile can be subharmonic at large depth.  Although
studying the effect of the subharmonic motion on the surface waves can
be interesting, here we limit ourselves to the simpler case of no such
motion.  The minimum depth at which the subharmonic motion occurs
depends on the interaction parameters, and is an increasing function
of $k_{n}$.  We thus use a rather large value of $k_{n} (10^{7})$, so
the subharnomic motion is absent in all cases studied here.  As a
check, we study the motion of the particles in a narrow tube of $W=1$
(five particle width) for several $\Gamma$ with fixed $f$.  Here,
$\Gamma$ is dimensionless peak acceleration $4 \pi A f^{2} / g$.  We
find that the pile behaves like a single particle for at least $10$
particle depth.  In particular, a bifurcation of the motion occurs at
$\Gamma \simeq 3.7$, which then terminates at $\Gamma \simeq 4.4$,
which agree well with the results of a single inelastic particle
\cite{gh83}.


We proceed to study surface waves.  We choose $W = 16$ and $N = 800$.
Thus the system is, on average, $80$ particle wide and $10$ particle
deep.  We fix the frequency $f = 10$, and study the waves for several
values of $\Gamma$.  We find that $f/2$ waves start to appear for
$\Gamma \sim 2.5$, then disappear when $\Gamma$ becomes about $4$.
When $\Gamma$ further increases, $f/4$ waves start to appear for
$\Gamma \sim 5.5$.  The features of this ``phase diagram'' agree well
with those of the experiments \cite{w97}.


Inspection of the motion of the particles shows that horizontal
movements of the particles play a crucial role in maintaining the
waves---an observation which was made in many previous studies on the
problem (e.g, \cite{ums98}).  We focus on the horizontal movements
which we characterize by average horizontal {\em speed} $U(t)$,
defined as
\begin{equation}
U(t) = {1 \over N} \sum_{i=1}^{N} | v_{x}^{i}(t) |,
\end{equation}
where $v_{x}^{i}$ is the horizontal velocity of particle $i$.  Even if
there are active horizontal movements, the average horizontal {\em
velocity} can be small since the movements can occur in both positive
and negative $x$ directions.  We thus use $U(t)$ instead of the
average horizontal velocity.  We normalize $U(t)$ with $2 \pi A f$,
the maximum velocity of the bottom plate.  In Fig.~1, we show
normalized $U(t)$ for $\Gamma = 2$ and $3$ for $10$ vibration periods.
Here, $f = 10$, $W = 16$, and $N = 800$.  The $\Gamma = 3$ curve has
been offset for clarity.

\begin{figure}[ht]
\centering
\includegraphics[angle=270,width=0.45\textwidth]{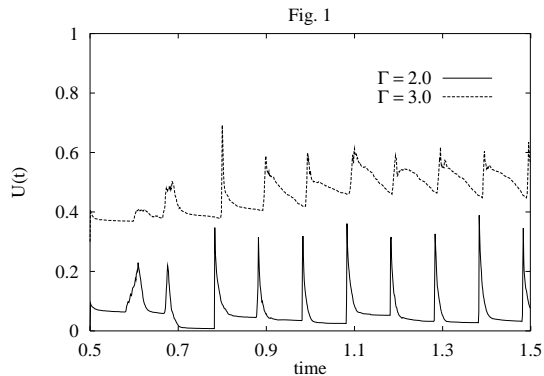}
\caption{Dimensionless average horizontal speed $U(t)$ for first $10$
vibration cycles.  Here, $W = 16$, $N = 800$, $f = 10$ and $\Gamma =
2$ and $3$.  The $\Gamma = 3$ curve has been offset for clarity.}
\label{fig:fig1}
\end{figure}

As shown in the figure, there clearly exist two separate processes:
sharp increase in $U$ within short time intervals and gradual decay of
$U$ during other phases of the vibration.  We also measure the time
series of the total pressure on the plate $p(t)$, which consists of
sharp peaks occurring when the pile collides with the plate.  We find
that the locations of the peaks in $U(t)$ coincide with those in
$p(t)$.  Thus, the horizontal movements are being fed by collisions of
the pile with the plate, and are being lost while the pile is not in
contact with the plate.

In order to maintain the surface waves, the particles have to travel
distance $\lambda$ within the period of the surface waves, where
$\lambda$ is their wavelength.  When $U(t)$ decays slowly, we expect
that the particles can travel long enough distance to maintain the
waves.  Indeed, we find that the surface waves are present only when
$U(t)$ decays slowly.  To be more precise, the waves are observed when
the decay time $\tau$, which will be defined later, is comparable to
the period of the vibration.

These observations strongly support the theories \cite{cmr97,s97,er98}
which argue that the horizontal motion is being supplied by collisions
with the bottom plate, and is being dissipated by interparticle
collisions.  The short time intervals during which the pile is
colliding with the plate play a crucial role in maintaining the
surface waves.


\begin{figure}[ht]
\centering
\includegraphics[angle=270,width=0.45\textwidth]{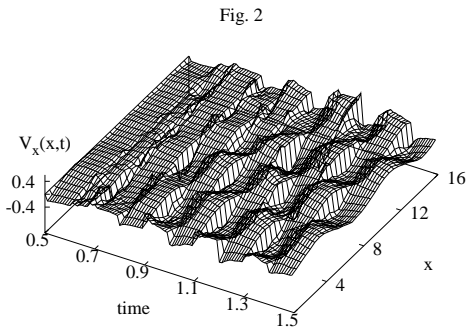}
\caption{Time evolution of the dimensionless horizontal velocity field
$V_{x}(x,t)$.  Here $\Gamma = 3$, $f = 10$, and other parameters are
the same as in Fig.~1.  The sudden increases in $V_{x}(x,t)$ occur
when the pile collides with the plate.}
\label{fig:fig2}
\end{figure}

We then consider each process in more detail.  In Fig.~2, we show time
evolution of the horizontal velocity field $V_{x}(x,t)$, which is
defined as the average horizontal velocity of the particles whose
centers are in $[x,x+dx]$.  Here, $\Gamma = 3$, $f = 10$, and other
parameters are the same as in Fig.~1.  We also normalize $V_{x}(x,t)$
by $2 \pi A f$.  In the figure, sharp changes in $V_{x}(x,t)$ occur at
certain values of $t$, which we find to be identical to the positions
of the peaks in $U(t)$.  Thus, sudden increases in $V_{x}(x,t)$ occur
when the pile collides with the plate.  The shape of $V_{x}(x,t)$
after such collisions does not change much until next collisions, but
the overall magnitude of $V_{x}(x,t)$ decreases.

The profile of $V_{x}$ just after the collisions is important for the
dynamics of the surface waves.  Since the motion of the particles is
deterministic, it is possible that the profile is determined from
certain field just before the collisions.  We check the dependency of
$V_{x}$ on several fields.  Specifically, we calculate the correlation
coefficient $r$ between $V_{x}$ after the collisions and the following
fields just before the collisions: $x$ and $y$ velocity fields $V_{x}$
and $V_{y}$, their spatial derivative $\partial _{x} V_{x}$ and
$\partial _{x} V_{y}$, the number density $m$ and its derivative
$\partial _{x} m$, and the center of mass field $h$ and its derivative
$\partial _{x} h$.  Here, $m(x)dx$ ($h(x)$) is defined to be the total
number (the center of mass) of the particles whose centers are in
$[x,x+dx]$.  Since the determination of the collision times becomes
difficult for large $\Gamma$, we study only for $\Gamma \le 4$.  We
average $r$ over $10$ collisions.  We find that $r$ is large ($\sim
-0.8$) for $\partial _{x} m, \partial _{x} h$ and $V_{x}$, and there
is essentially no correlation with the other fields.  The large value
of $r$ for $\partial _{x} h$ suggests
\begin{equation}
\label{eq:flux}
V_{x}(x,t_{a}) \propto - \partial _{x} h(x,t_{b}),
\end{equation}
where $t_{a}$ and $t_{b}$ is the time just after and before the
collisions, respectively.  This very equation was assumed in the
theories of Cerda {\it et al.} \cite{cmr97} and Eggers and Riecke
\cite{er98}, where they proposed the form motivated from the flow of
granular material on an inclined plane.  The strong correlation occurs
for all $\Gamma$ studied here, including the $\Gamma = 2$ case where
no surface wave is observed.

We find that the spatial variation of the packing fraction is not
significant, so $m$ is roughly proportional to $h$.  We thus expect
that the correlation for $\partial _{x} m$ is also large.  What is
strange is the strong correlation between $V_{x}$ before and after the
collisions, which states that the particles reverse the direction of
their horizontal movements while the pile is colliding with the plate.
We do not understand the origin of the reversal of the motion.

We also measure the magnitude of the horizontal movements for several
values of $\Gamma$.  To be more specific, we measure the peak values
of $U(t)$ like the ones shown in Fig.~1, and averaged them over all
the collisions.  One might expect that the average is proportional to
the collision velocity of particles on the plate.  Even though both
display minimum around $\Gamma = 4.6$, we find that there is no strong
correlation between them, and the average seems to show complicated
dependence on $\Gamma$.  Further study is necessary to quantify and
understand the dependence.


Next, we study the decay process, where horizontal movements of the
particles gradually decrease between the collisions with the plate.
We quantify the process as follows.  We start from a time series of
$U(t)$ like the ones in Fig.~1.  We translate the positions of the
peaks so that they all coincide.  We then normalize the peak values of
$U(t)$ to unity, and average $U(t)$ over all the peaks.  The resulting
quantity, defined as $U_{a}(t)$, can be used to characterize the decay
process.  In Fig.~3, we show $U_{a}(t)$ for $\Gamma = 2, 3$ and $4$,
where other parameters remain unchanged.  We then define decay time
$\tau$ as the ``half-life''---the time at which $U_{a}(t)$ becomes
$1/2$.  The half-life for $\Gamma = 2$ is about $0.0045$, which is
much smaller than the period of the vibration ($T = 0.1$).  On the
other hand, $\tau$ is close to $0.08$ for $\Gamma = 3$ and $4$.  For
all $\Gamma$ we have studied, we find that the surface waves are
present if and only if $\tau$ is comparable to the period of the
vibration.  The need for large $\tau$ for the formation of the surface
waves is evident as previously discussed.  The fact that the dynamics
consists of two separate processes does not seem to depend on small
variations of interaction parameters.

\begin{figure}[ht]
\centering
\includegraphics[angle=270,width=0.45\textwidth]{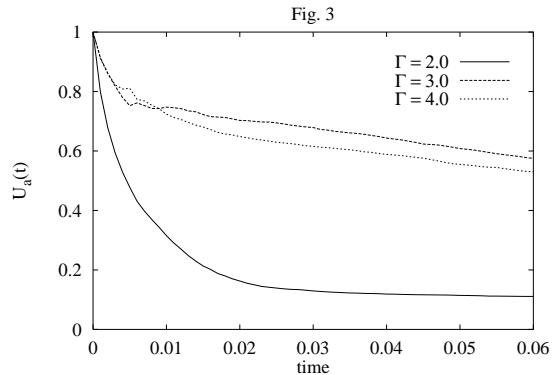}
\caption{The time evolution of $U_{a}(t)$ for $\Gamma = 2, 3$ and $4$.
The ``half-life'' $\tau$ is small for $\Gamma = 2$, but becomes
comparable to the period of the vibration for $\Gamma = 3$ and $4$.}
\label{fig:fig3}
\end{figure}


The quantitative understanding of the decay process requires that one
should be able to predict the dependence of $\tau$ on parameters such
as $\Gamma$.  However, it seems that such dependences are rather
complex.  For example, we find that $\tau$ does not depends on
$\Gamma$ in a simple fashion: $\tau$ is not a monotonic function of
$\Gamma$, and a small increase in $\Gamma$ can result in a ten-fold
increase or decrease in $\tau$.  At the particle level, the decay
process occurs by the collisions between the particles, and the
resulting changes in the horizontal movements.  In order to make
quantitative predictions, one thus has to understand both the
collision frequency and the resulting momentum changes.
Unfortunately, both of these quantities are poorly understood.  More
studies are again needed in order to understand the decay process.


The time evolution of horizontal movements of the particles can also
be studied using auto-correlation function $c(t)$ defined as $\langle
v_{x}(0) v_{x}(t) \rangle$, where the average is taken over the
particles.  We find that $c(t)$ also displays two basic processes:
sharp change during the collisions with the plate and slow decay
between them.


In sum, we show that horizontal movements of the particles, which are
essential to the formation of the surface waves, consist of two
separate processes: sharp increase of the movements by the collisions
with the bottom plate, and slow decrease of the movements due to
interparticle collisions, which strongly support the theories of
\cite{cmr97,s97,er98}.  We also find that the horizontal velocity
field $V_{x}(x)$ after the collisions with the plate is strongly
correlated to $\partial _{x} h(x)$ just before the collisions.  Here,
we are mainly interested in qualitative features of the mechanism for
the formation of the surface waves.  Their quantitative understanding,
such as the parameter dependence of $\tau$, will be subject of future
works.

After the present work has been completed, we became aware of the work
of Kim {\it et al} \cite{kpc98}, where similar results were obtained.


I thank H. K. Pak, S. Kim, K. Kim and S.-O. Jeong for useful
discussions.  This work is supported in part by the Department of
Energy under grant DE-FG02-93-ER14327, Korea Science and Engineering
Foundation through the Brain-Pool program, and SNU-CTP.

\end{document}